\documentclass[%
reprint,
10pt,
superscriptaddress,
nofootinbib,
 amsmath,amssymb,
 aps,
 prl,
 twocolumn,
]{revtex4}

\usepackage{graphicx}
\usepackage{dcolumn}
\usepackage{bm}
\usepackage[colorlinks=true,urlcolor=blue,citecolor=blue,linkcolor=blue]{hyperref}

\usepackage{physics}     
\usepackage{xcolor}      
\usepackage{color,soul}
\usepackage{hyperref}    
\usepackage{float}       
\usepackage{transparent} 
\usepackage{layouts}     

\begin{document}
\title{Thermalization of quantum light induced by classical nonlinear wave dynamics}
\author{Fouad Chahrour}
\email[Electronic address: ]{fouad.chahrour@physik.hu-berlin.de}
\affiliation{Max-Born-Institut, Max-Born-Strasse 2A, 12489 Berlin, Germany}
\affiliation{Humboldt-Universität zu Berlin, Institut für Physik, AG Theoretische Optik \& Photonik, Berlin, Germany}
\author{Şahin K. Özdemir}
\affiliation{Department of Electrical and Computer Engineering, Saint Louis University, St. Louis, MO 63103, USA}
\author{Kurt Busch}
\affiliation{Max-Born-Institut, Max-Born-Strasse 2A, 12489 Berlin, Germany}
\affiliation{Humboldt-Universität zu Berlin, Institut für Physik, AG Theoretische Optik \& Photonik, Berlin, Germany}
\author{Ramy El-Ganainy}
\email[Electronic address: ]{relganainy@slu.edu}
\affiliation{Department of Electrical and Computer Engineering, Saint Louis University, 
St. Louis, MO 63103, USA}
\author{Armando Perez-Leija}
\email[Electronic address: ]{armando.leija@slu.edu}
\affiliation{Department of Electrical and Computer Engineering, Saint Louis University, St. Louis, MO 63103, USA}
\date{\today}

\begin{abstract}

Thermalization of isolated quantum systems is an intriguing phenomenon at the forefront of contemporary physics. In this work, we demonstrate that nonlinear multimode optical platforms can be harnessed to induce effective quantum interactions between photons. Through numerical experiments where quantum beams propagate alongside classical light within multimode nonlinear optical systems, we reveal the thermalization of fundamental quantum light states—specifically single- and two-photon states. This thermalization is clearly manifested by the emergence of Rayleigh-Jeans and Boltzmann statistical distributions. Beyond providing a deeper understanding of how classical nonlinearities can be used to investigate quantum many-body dynamics, our findings will enable the exploration of a broader range of complex quantum phenomena, including aspects of quantum phase transitions, within readily accessible classical optical settings.
\end{abstract}

\maketitle

\section{Introduction}
Thermalization is a fundamental relaxation process through which dynamically interacting systems evolve toward states where macroscopic thermodynamic quantities become stationary~\cite{PolkovnikovRevModPhys}. From a microscopic perspective, this corresponds to the maximization of entropy~\cite{LindenPRE}. Thermalization emerges from the interplay between two competing physical processes: a reversible one, such as Hamiltonian (unitary) evolution, and an irreversible one. In open systems, irreversibility arises naturally through interactions with an infinitely large reservoir. In contrast, closed systems can exhibit thermalization due to nonlinear interactions that are highly sensitive to initial conditions. This is a hallmark of chaos, which renders the dynamics effectively irreversible, despite the underlying equations of motion being time reversible~\cite{StrogatzBook}. Although these concepts are well established in classical physics, the thermalization of quantum systems remains a very active field of research~\cite{Rigol2008,RigolPRL2012,GoldsteinPRL2015,BrenesPRLett2020,Vidmar2016}. This is primarily because many classically nonlinear systems are governed, at the quantum level, by linear equations of motion. The apparent nonlinearity in the classical regime often emerges only through approximations such as mean-field factorization~\cite{ChaikinBook}.

Recently, the notion of optical thermodynamics has emerged as a powerful framework for investigating light dynamics in multimode nonlinear optical systems \cite{Wu2019_Thermodynamic}. Although concepts from statistical mechanics have previously been employed to study nonlinear optical interactions \cite{Aschieri2011_Condensation,Picozzi2012_Condensation,LevyPRB2018,LevyJoPB2018,SilberbergPRL2009}, the work in \cite{Wu2019_Thermodynamic} was the first to establish a clear connection to classical thermodynamics, applicable across a broad class of multimode optical systems. In classical optics, thermalization manifests in the mode distributions of nonlinear systems. Indeed, such nonlinearity-induced optical thermalization has been experimentally observed in a variety of multimode platforms, including optical fibers \cite{Pourbeyram2022_Direct} and time-synthetic photonic lattices \cite{Marques2023_Observation}. These studies have revealed a universal trend: the entropy of multimode light states propagating through nonlinear optical media tends to increase, leading to steady states whose modal populations are well described by the Rayleigh-Jeans distribution \cite{Parto2019_Thermodynamic}. Although initial investigations focused on Kerr-type nonlinearities, it was soon demonstrated that optical thermalization is a robust and generic process that arises in a wide range of nonlinear interactions, including non-perturbative regimes \cite{Zhong2023_Universality}.

Importantly, modal thermalization in classical optical systems requires only weak nonlinear interactions. In fact, weak nonlinearity is a necessary condition to avoid soliton formation, which would otherwise inhibit thermalization by localizing energy in coherent non-dispersive structures. This raises an important question: Can similar systems be utilized to explore thermalization in the quantum regime, where particle statistics—bosonic versus fermionic—can play a crucial role? However, if the nonlinear interaction is too weak, the timescales (or propagation distances, in the case of waveguides) required to reach thermal equilibrium become prohibitively long, rendering experimental observations impractical. This creates a fundamental challenge: the intermediate nonlinear regime—strong enough to induce thermalization within realistic experimental constraints, yet weak enough to avoid coherent structures like solitons—remains elusive in quantum optical systems. Numerous studies have explored few-photon-level interactions \cite{Peyronel2012,Chang-Lukin2014}. However, none have yet achieved the strong and controllable nonlinearity to enable quantum thermalization under accessible laboratory conditions.\\

\begin{figure*}[t!]
\centering
\includegraphics[width=0.98\textwidth]{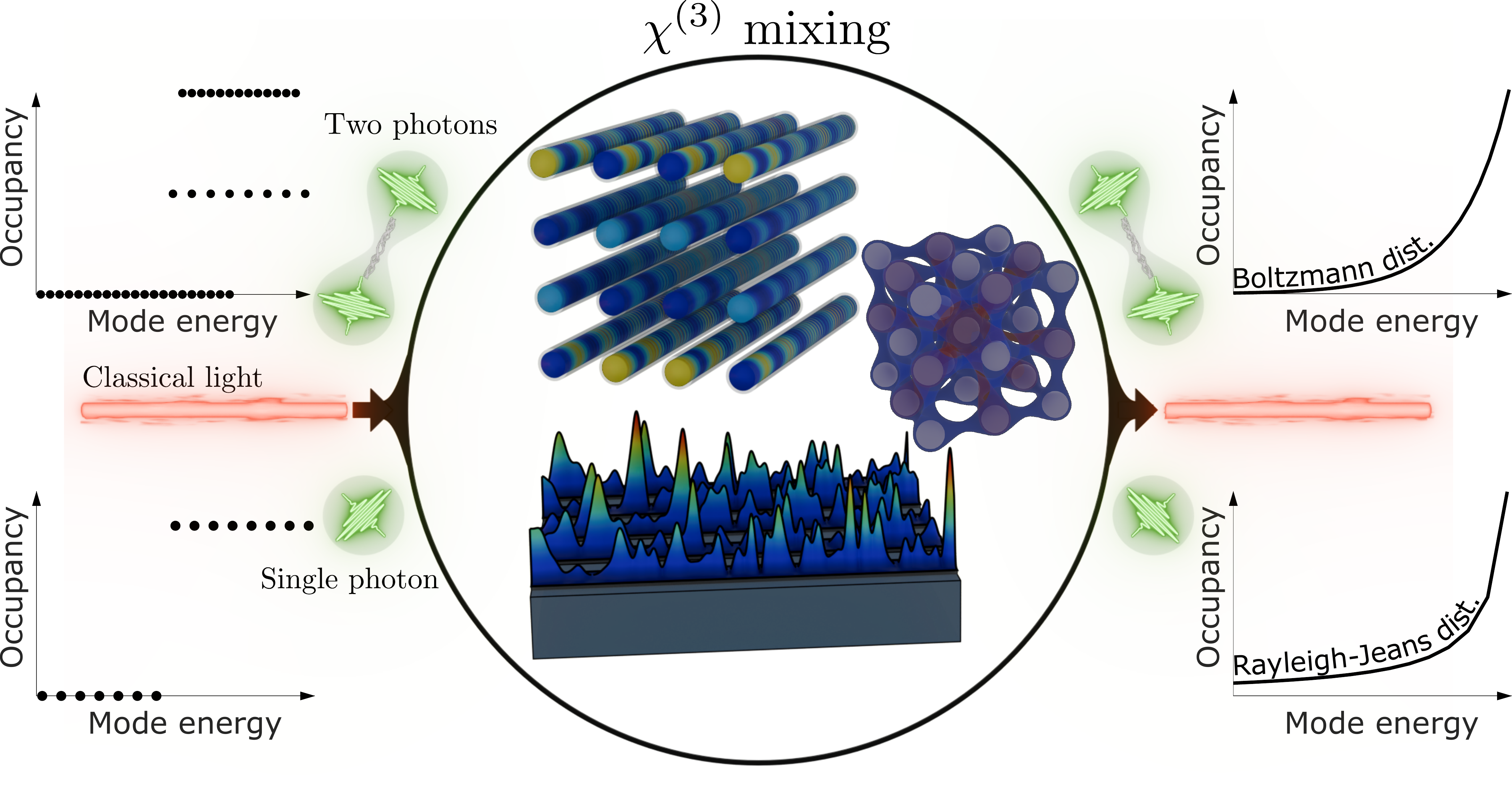}
\caption[]{\label{fig:schematicapproach3} Conceptual sketch of nonlinearity-induced thermalization of quantum light. (Left) Classical and quantum light simultaneously injected into nonlinear multimode coupled systems. Each field enters the system with a well defined mode distribution or mode occupancy. (Center) Nonlinear evolution of classical light induces chaotic fluctuations in the effective refractive index. (Right) Boltzmann and Rayleigh-Jeans output mode occupancies for two-photon and single-photon light, respectively.}
\label{fig:Scheme}
\end{figure*}

In this work, we propose a scheme to overcome the challenges of quantum optical thermalization by demonstrating that interactions between classical and quantum light in nonlinear waveguide platforms can drive thermalization of quantum states. Our approach, illustrated schematically in Fig.~(\ref{fig:schematicapproach3}), relies on a Kerr-type nonlinear coupling between a classical light field and a quantum state of light (such as single photons or correlated photon pairs) with orthogonal polarizations (or different frequencies), ensuring their mutual distinguishability. In this configuration, each polarization component experiences both self-phase modulation and cross-phase modulation. However, due to the low intensity of the quantum light, its self-interaction is negligible, and the classical field evolves independently under its own nonlinear dynamics. In contrast, the quantum field primarily experiences cross-phase modulation, governed by the fluctuations in the intensity profile of the classical component. This separation of dynamics allows us to first compute the evolution of the classical field and then treat its intensity distribution along the propagation direction as a background potential driving the unitary evolution of the quantum state. Remarkably, we find that the classical field's intensity profile becomes effectively chaotic due to nonlinear self-interactions, leading to thermalization of the quantum light. 
Crucially, our results show that the final thermal state depends on the photon statistics of the input quantum state: the steady-state distribution resulting from two-photon states differs from that of a single-photon input. This highlights the central role of particle statistics in the thermalization process, even when mediated by classical nonlinear dynamics.\\
\section{Results}
\noindent
\textbf{System and model.} 
In this work, we consider a system composed of coupled optical waveguides. While we start primarily
dealing with waveguide arrangements in order to exemplify
our approach, a similar discussion will be shown for a three-dimensional multimode cavity array.
In the waveguide system, each guide supports one transverse electric (TE) and one transverse magnetic (TM) mode. Evanescent coupling between adjacent waveguides allows both TE and TM modes to experience discrete diffraction across the array. Further, the material is assumed to exhibit Kerr nonlinearity, with nonvanishing tensor components that introduce cross-polarization nonlinear interactions. Under these conditions, the quantum Hamiltonian of the system is
\begin{equation}
\begin{split}
\mathcal{H} &= \sum_n \hbar \epsilon_n^{(a)} \hat{a}^{\dagger}_n \hat{a}_n + \sum_n \hbar \epsilon_n^{(b)} \hat{b}^{\dagger}_n \hat{b}_n \\
&+ \sum_{n \neq m} \hbar \kappa_{nm}^{(a)} \hat{a}_n^\dagger \hat{a}_m + \sum_{n \neq m} \hbar \kappa_{nm}^{(b)} \hat{b}_n^\dagger \hat{b}_m \\
&+ \sum_n \hbar S \hat{a}_n^{\dagger} \hat{a}_n \hat{a}_n^{\dagger} \hat{a}_n + \sum_n \hbar S \hat{b}_n^{\dagger} \hat{b}_n \hat{b}_n^{\dagger} \hat{b}_n \\
&+ \sum_n \hbar C \hat{a}_n^{\dagger} \hat{a}_n \hat{b}_n^{\dagger} \hat{b}_n + \text{H.c.}
\end{split}
\end{equation}
where $\hat{a}_n^{\dagger}$ and $\hat{a}_n$ are the creation and annihilation operators associated with one polarization mode (e.g., TE), while $\hat{b}_n^{\dagger}$ and $\hat{b}_n$ are associated with the orthogonal polarization (e.g., TM) at waveguide $n$ (or cavity $n$). The parameters $\epsilon_n^{(a)}$ and $\epsilon_n^{(b)}$ are the respective propagation constants (cavity resonance frequencies), and $\kappa_{nm}^{(a)}$, $\kappa_{nm}^{(b)}$ represent the evanescent coupling strengths between neighboring waveguides for each polarization. The coefficients $S$ and $C$ represent the strengths of the nonlinear self-phase and cross-phase interactions, respectively, while $\text{H.c.}$ denotes the Hermitian conjugate.\\

From the above Hamiltonian, we can derive the Heisenberg equations of motion for the operators $\hat{a}_n$ and $\hat{b}_n$. To proceed, we treat mode $a$ as a classical field by replacing the operator with its expectation value, $\langle \hat{a}_n \rangle = \alpha_n$, and neglect the influence of mode $b$ on mode $a$ (justified by the low intensity of the quantum field in mode $b$). Furthermore, we ignore the self-phase modulation of mode $b$. Under these approximations, the evolution equations for the co-propagating classical and quantum components acquire the form
\begin{subequations} \label{eq:eom}
\begin{align}
i\frac{d}{dz} \ket{\boldsymbol{\alpha}(z)} &= \left(\mathbf{H}^a_\mathrm{L} + \mathbf{H}_\mathrm{NL}^{a}(z)\right) \ket{\boldsymbol{\alpha}(z)}, \label{eq:eom_a} \\
i\frac{d}{dz} \hat{B}(z) &= \left(\mathbf{H}_\mathrm{L}^{b} + \mathbf{H}_\mathrm{NL}^{b}(z)\right) \hat{B}(z). \label{eq:eom_b}
\end{align}
\end{subequations} 
Here, $\ket{\boldsymbol{\alpha}} = [\alpha_1(z), \alpha_2(z), \ldots]^T$ represents the classical field vector, and $\hat{B}(z) = [\hat{b}_1(z), \hat{b}_2 (z), \ldots]^T$. The matrices $\mathbf{H}_\mathrm{L}^{a,b}$ govern the linear dynamics of modes $a$ and $b$, incorporating both propagation constants $\epsilon_n^{(a,b)}$ and coupling coefficients $\kappa_{nm}^{(a,b)}$. Concurrently, the nonlinear matrices $\mathbf{H}_\mathrm{NL}^{a,b}(z)$ capture the self and cross-phase modulation induced by the classical field and are both diagonal, with entries proportional to $S |\alpha_n(z)|^2$, and $C |\alpha_n(z)|^2$, respectively. 
To solve this system of equations, we first transform the classical field according to $\ket{\boldsymbol{\alpha}} \rightarrow \ket{\boldsymbol{\alpha}}/\rho$, where $\rho=1/\sqrt{2S}$, and integrate Eq.~\eqref{eq:eom_a} numerically. 
The resulting classical intensity profile is then used to construct $\mathbf{H}_\mathrm{NL}^{b}(z)$ for substitution into Eq.~\eqref{eq:eom_b}, which is subsequently solved numerically. Note Eq.~\eqref{eq:eom_b} is a linear, $z$-dependent (or time-dependent in the case of cavities) equation, with a formal solution given by the unitary operator
$
\hat{\mathcal{U}}(z, z_0) = \mathcal{Z} \exp\left(-i \int_{z_0}^{z} H_{\text{tot}}(z')\, dz'\right)$
where \( H_{\text{tot}}(z) = \mathbf{H}_\mathrm{L}^{(b)} + \mathbf{H}_\mathrm{NL}^{(b)}(z) \), and \( \mathcal{Z} \) denotes the $z$-ordering operator, analogous to the time-ordering operator in time-dependent quantum mechanics.
 
In what follows, and without loss of generality, we assume \( \mathbf{H}_\mathrm{L}^{(a)} = \mathbf{H}_\mathrm{L}^{(b)} \equiv \mathbf{H}_\mathrm{L} \) and \( S = C \). 
Although the actual values of these parameters are determined by the linear properties of the modes and the nonlinear characteristics of the underlying material system, their precise values are not critical for the qualitative analysis presented here. 
For clarity, we therefore omit the superscripts from the expressions for both the linear and nonlinear Hamiltonians.\\

Before we proceed further, it is important to summarize the basic concepts of nonlinear optical thermodynamics~\cite{Wu2019_Thermodynamic,Parto2019_Thermodynamic,Zhong2023_Universality}, which predict the spatial thermalization of the classical modes $a$ governed by Eq.~(\ref{eq:eom_a}). In the absence of dissipation, the optical power of the system, given by $P \equiv \braket{\boldsymbol{\alpha}}{\boldsymbol{\alpha}} = \sum_{i=1}^{M} |c_i|^2$, is conserved. In this expression, $c_i = \braket{\phi_i}{\boldsymbol{\alpha}(0)}$, where $\phi_i$ is the eigenvector of $\mathbf{H}_\mathrm{L}$ with eigenvalue $\epsilon_i$. Under weak nonlinearity, we can define the internal energy $U \equiv -\bra{\boldsymbol{\alpha}(0)} \mathbf{H}_\mathrm{L} \ket{\boldsymbol{\alpha}(0)} = -\sum_{i=1}^{M} \epsilon_i |c_i|^2$, where the negative sign reflects the inherent inversion in the spectrum of optical potentials~\cite{Parto2019_Thermodynamic}. Maximizing the entropy under these conditions and the assumption of fix average number of photons
yields the RJ distribution for mode occupancies, $|c_i|^2 = -T/\left(\epsilon_i + \mu\right)$, where $\mu$ is the optical analogue of the chemical potential and $T$ is the optical temperature~\cite{Wu2019_Thermodynamic,Parto2019_Thermodynamic}. Using this distribution, the internal energy and power become $U = T \sum_{i=1}^{M} \frac{\epsilon_i}{\epsilon_i + \mu}$ and $P = -T \sum_{i=1}^{M} \frac{1}{\epsilon_i + \mu}$, which together yield the equation of state $U - \mu P = M T$. Therefore, once the set of $M$ eigenvalues is known, the optical temperature of a nonlinear multimode system can be uniquely determined for any given internal energy $U$ and power $P$. This nonlinearity-induced RJ distribution has been experimentally observed in the ensemble-averaged modal responses of multimode nonlinear optical systems excited with random-phase multimode fields~\cite{Pourbeyram2022_Direct}. Throughout this work, we exploit the ergodicity of the system and generate combined averages over the ensemble and over time (propagation distance).

In the present work, we consider three nonlinear multimode systems: i) the so-called $J_x$ lattice~\cite{Perez2013_Coherent,Perez2013_Perfect,Tschernig2018Multiphoton}, ii) a two-dimensional waveguide array, and iii) a three-dimensional cavity array. 
For the $J_x$ system, we assume an array composed of $N=15$ waveguides with $\kappa_m = \frac{1}{2}\sqrt{(N - m)m}$ representing the coupling strength between waveguide elements $m$ and $(m+1)$. We emphasize that there is nothing inherently special about $J_x$ arrays—rather, their ladder-like spectrum leads to faster thermalization, which in turn facilitates numerical integration. Assuming an input classical field composed by the mode distribution illustrated in Fig.~(\ref{fig:thermalizationsinglephoton}a) results in a power $P = 1$ and internal energy $U = -3.5$. Under these initial conditions, the optical thermodynamic model~\cite{Parto2019_Thermodynamic} predicts a steady-state response consisting of an incoherent superposition of supermodes, with weights given by the RJ distribution at temperature $T = 0.3$ and chemical potential $\mu = -7.98$, see continuous curve in Fig.~(\ref{fig:thermalizationsinglephoton}a). Numerical integration of Eq.~(\ref{eq:eom_a}) for this initial condition over a normalized propagation distance $z=5\times 10^4$, and averaging over $50$ numerical realizations and the last $20\%$ of the numerically computed solutions produces the mode distribution described by the dashed (red-crosses) curve in Fig.~(\ref{fig:thermalizationsinglephoton}a), which is in good agreement with the theoretically predicted distribution. 
Figures~(\ref{fig:thermalizationsinglephoton}b) and ~(\ref{fig:thermalizationsinglephoton}c) show the results for the 2D waveguide array and the 3D cavity array.\\

\begin{figure*}[t]
    \includegraphics[width=0.99\textwidth]{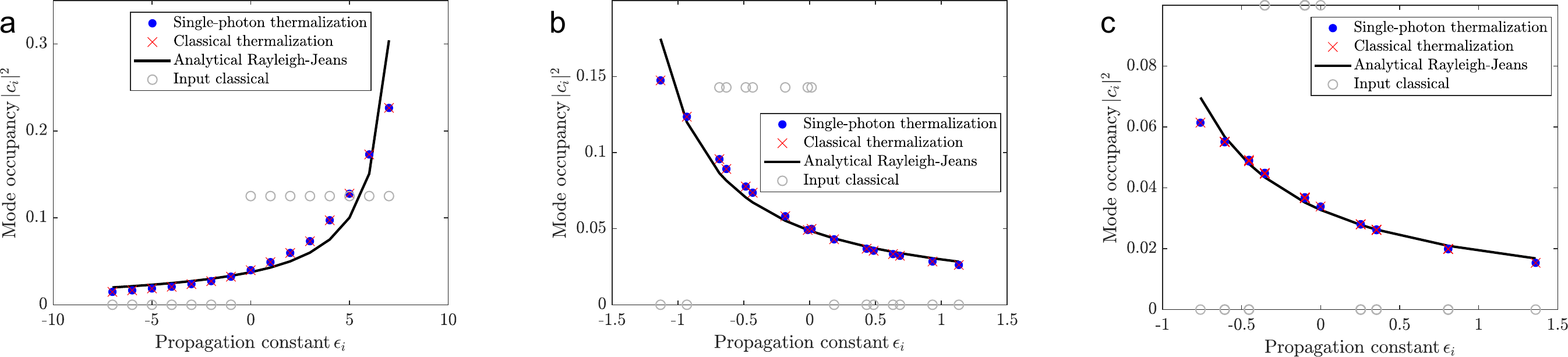}
    \label{fig:thermalizationsinglephoton}
	\caption[]{\label{fig:thermalizationsinglephoton}
	Thermalization of classical and single-photon light beams in nonlinear waveguides and a 3D cavity array. \textbf{a} Nonlinear classical and single-photon thermalization computed for a $J_x$ array after a propagation distance $z=5\times 10^4$ and the analytically computed RJ distribution predicting a temperature $T=0.3$ and chemical potential $\mu=-7.98$. \textbf{b} The same as in \textbf{a} for a nonlinear two-dimensional waveguide array after a propagation distance $z=1.5 \times 10^6$. The corresponding analytical RJ distribution is predicted to have temperature $T=-0.08$ and chemical potential $\mu=1.57$. \textbf{c} Nonlinear classical and single-photon thermalization computed for a three-dimensional cavity system after a propagation time $z=8\times 10^5$ and the corresponding RJ distribution with temperature $T=0.05$ and chemical potential $\mu=-1.44$. In all three cases the continuous line represents the analytically obtained RJ distribution, the dot-dashed and cross-dashed lines correspond to the numerically obtained single-photon and classical light distributions, respectively.}
\end{figure*}

\noindent
\textbf{Single-photon thermalization.}
We now examine the thermalization of single-photon states. Specifically, we consider Eqs.~(\ref{eq:eom}) for input states of the form
\begin{equation}
    \ket{\boldsymbol{\alpha}(0)} = \sum_n c_n(0) \ket{v_n}, \qquad 
    \ket{\Psi(0)} = \sum_n d_n(0) \hat{w}_n^{\dagger} \ket{0_{\text{S}}}.
\end{equation}
Here, $\ket{v_n}$ are the eigenvectors of the Hamiltonian $\mathbf{H}_\mathrm{L}$, and $\hat{w}_n^{\dagger}$ are the bosonic creation operators in the supermode basis. That is, $\hat{w}_n^{\dagger} \ket{0_{\text{S}}}$ represents a single photon populating supermode $\ket{v_n}$. The subscript ``S'' emphasizes that the vacuum state $\ket{0_{\text{S}}}$ is defined in the supermode basis. From our previous discussion, we know that the distribution of the classical coefficients $c_n(z_f)$---more precisely, their squared magnitudes $|c_n(z_f)|^2$---obeys the RJ distribution at the output plane $z = z_f$, see dashed (red-crosses) curves in Figs.~(\ref{fig:thermalizationsinglephoton} a,b,c).\\
\indent
We now use the intensities associated to the numerical field vectors that give rise to the classical thermal mode distributions for the construction of the corresponding $\mathbf{H}_\mathrm{NL}^{ab}(z)$. Then, we numerically integrate Eq.~(\ref{eq:eom_b}) to establish the input-output relations for the operators $\hat{B}^\dagger(z_f)$ and $\hat{B}^\dagger(0)$. These relations allow us to determine the coefficients $d_n(z_f)$ and, consequently, the thermal distribution $|d_n(z_f)|^2$, which characterizes the mode occupation probabilities of single-photon states, see Fig.~(\ref{fig:thermalizationsinglephoton}).
Clearly, we find that the single-photon state follows exactly the same mode distribution as the classical field, and this occurs despite the fact Eq.~(\ref{eq:eom_b}) is linear. The key feature that enables the thermalization of single-photon states is the effective chaotic nature of the $z$-dependent classical intensity distributions in the waveguides, which correspond to the diagonal entries in matrix $\mathbf{H}_\mathrm{L}(z)$.

Figures~(\ref{fig:thermalizationsinglephoton} b, c) the corresponding results for the 2D waveguide array and the 3D cavity array. In both cases, 
we observe thermalization of the classical fields and single-photon states to the RJ distribution, with good agreement between numerical integration and theoretical predictions.\\
\indent
Up to this point, we have tacitly assumed input classical fields with power $P=1$, such that they match the total probability $(\mathcal{P}=1)$ in the normalized single-photon wavefunctions. This premise conditioned the single photons to evolve toward thermal steady states described by RJ distributions identical to the ones exhibited by the corresponding classical fields. However, the observation of classical mode thermalization is only constrained by the need of power conditions that induce weak nonlinear interactions, as discussed in the Introduction.\\ 
\indent
To explore more general scenarios, we performed numerical experiments assuming classical input fields with powers $P>1$. Interestingly, these simulations reveal that both classical and single-photon fields exhibit mode distributions with identical shapes, differing only by a scaling factor. However, normalizing the classical RJ distributions results in a perfect match with the distributions computed for single photons.
These findings strongly suggest that the underlying classical thermalization mechanism, that is, the nature of the nonlinear mode-mixing interactions, has been identically imprinted onto the single-photon dynamics.

These effects can be explained by expressing the single-photon fields in terms of the eigenmodes of $\mathbf{H_L}$, $\ket{\Psi(z)}=\sum_{m}c_{m}(z)\ket{\varphi_{m}}$, and substituting into the Schrödinger representation of equation Eq.~(\ref{eq:eom}b). This yields the equations of motion for the mode amplitudes
\begin{equation}\label{eq:amps}
i\frac{d}{dz}c_{m}(z)=\epsilon_{m}c_{m}(z)+\sum_{l=1}^{M}\kappa_{ml}(z)c_{l}(z),
\end{equation}
where $\kappa_{ml}(z)=\bra{\varphi_{m}}\mathbf{H}_\mathrm{NL}^{b}(z)\ket{\varphi_{l}}$ represents the $z$-dependent coupling between single-photon eigenmodes.
That is, the chaotic fluctuations observed in the intensity-dependent refractive index, which are encoded in the intensity matrix $\mathbf{H}_\mathrm{NL}^{b}(z)$, effectively induce chaotically fluctuating couplings among single-photon modes.\\

Before addressing the two-photon case, we examine the nature of the intensity fluctuations by computing the longitudinal intensity correlations in the nonlinear systems under consideration.
To do so, we compute the normalized intensity correlation function $g^{(2)}(\tau) = \langle I_{x}(z+\tau) I_{x}(z) \rangle/\langle I_{x}(z) \rangle^2$, where $\langle \cdot \rangle$ denotes the ensemble average and $x$ identifies the waveguide site over which the intensity is monitored. Notice fully coherent light is characterized by $g^{(2)}(0)=1$, whereas $g^{(2)}(0)=2$ witnesses fully chaotic thermal light \cite{Loudon2000_The}.

 The intensity correlations computed for our numerical experiments yield $g^{(2)}(0)\in(1.6,2)$, clearly revealing the competition between the reversible Hamiltonian dynamics and the chaotic irreversible nonlinear effects.\\ 

\noindent
\textbf{Two-photon thermalization.} On the one hand, the results presented in the previous section are intriguing, as they show that the effectively chaotic refractive index modulation induced by the classical beams drives a single photon to evolve
into a thermal state with a structure identical to the classical one. On the other hand, it is well known that single-photon states, although quantum in nature, can exhibit dynamics that are effectively indistinguishable from those of classical light \cite{Eberly2007}. It is therefore of interest to investigate how the introduction of quantum indistinguishability, specifically in two-photon states, modifies the thermalization dynamics observed in single-photon systems.\\
\indent
To address this question, we investigate whether the inherent quantum correlations and wavefunction symmetrization of two indistinguishable photons induce distinct thermalization dynamics within the supermode basis, potentially revealing unique quantum effects of multiparticle indistinguishability. To do so, we explore the dynamics resulting from co-propagating two-photon states through the multimode nonlinear optical systems in which the thermalization effects reported in Fig.~(\ref{fig:thermalizationsinglephoton}) take place. 
To reveal deviations from the established single-photon supermode thermalization pathway, as probe states we utilize coherent superpositions of two indistinguishable photons inhabiting input supermode distributions that involve the modes populated by the classical input classical fields.
In other words, we consider the following two-photon initial conditions and examine their evolution under the thermalizing influence of the classical beams

\begin{align}
	\label{eq:twophotoneigenmodes}
\ket{\Psi^{(2)}(0)}&=\sum_{m=1}^{M}c_{m,m} \ket{\phi_m}\ket{\phi_m} +\\
& \sum_{m=1}^{M}\sum_{n=m+1}^{M}c_{m,n} \frac{1}{\sqrt{2}} \left( \ket{\phi_m}\ket{\phi_n}
+ \ket{\phi_n}\ket{\phi_m}\right).\nonumber
\end{align}

\begin{figure*}[t!]
    \includegraphics[width=0.99\textwidth]{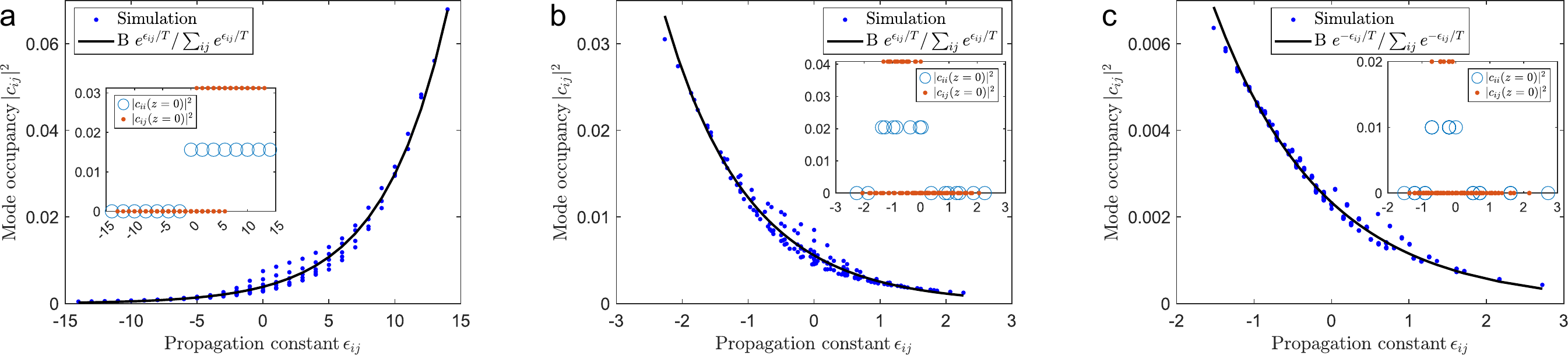}
	\caption[]{\label{fig:thermalizationtwophoton} Thermalization of two-photon light in nonlinear waveguide arrays and a cavity array. \textbf{a} Two-photon thermalization computed for a $J_x$ array after a propagation distance $z=5\times 10^4$ and corresponding analytically computed Boltzmann distribution with temperature $T=4.91$. \textbf{b} The same as in \textbf{a} for a two-dimensional waveguide array after a propagation distance $z=1.5\times 10^6$. The corresponding Boltzmann distribution is predicted to exhibit a temperature $T=-1.31$.distribution. \textbf{c} Two-photon thermalization corresponding to the three-dimensional cavity array after a propagation distance $z=8\times 10^5$ and characterized by a Boltzmann distribution with temperature $T=1.44$. In all three cases the insets depict the input two-photon mode distributions.}
\end{figure*}

In the above expression, the first summation corresponds to maximally entangled (non-separable) two-photon NOON states, in which both photons are simultaneously populating the same supermode, with equal probability across the \( M \) available modes. The second summation represents a coherent superposition of symmetrized states where one photon occupies mode \( m \) and the other mode \( n \). To avoid overcounting due to the symmetry of the two-photon wavefunction, the second sum in this term starts at \( n = m + 1 \). Hence, the initial state defined in Eq.~\eqref{eq:twophotoneigenmodes} comprises a total of \( M(M+1)/2 \) two-photon supermodes, each associated with an eigenvalue given by the corresponding combination of single-photon eigenvalues, \( \epsilon_{m,n} = \epsilon_m + \epsilon_n \). The mode occupancies for the initial states used in our simulations are illustrated in the inset panels of Figs.~(\ref{fig:thermalizationtwophoton} a-c).

The two-photon wavefunction at any propagation distance \( z \) is given by \( \ket{\Psi^{(2)}(z)} = U^{(2)}(z) \ket{\Psi^{(2)}(0)} \), where the two-photon propagator is expressed as the tensor product of single-photon propagators, \( U^{(2)}(z)=U(z)\otimes U(z) \). The corresponding mode occupation coefficients are given by \( |c_{ij}(z)|^2 = |\langle \phi^{(2)}_{ij} | \Psi^{(2)}(z) \rangle|^2 \).

We stress that similarly to the single-photon case, the power of the classical fields was set to $P=1$ to match the probability of the normalized two-photon wavefunctions. Furthermore, time averaging is performed on multiple simulations and the final 20\% of the simulated data to extract steady-state statistics. We display these results in Fig.~(\ref{fig:thermalizationtwophoton}).

To determine the equilibrium two-photon state, we compute the density operator that maximizes the entropy of the system, $S=-k_BTr\{\rho\ln(\rho)\}$, when it is subjected to the constraints $Tr\{\rho\}=1$, constant
average energy $\langle E \rangle$, and average photon number $\langle N \rangle$. By doing so, we find the density operator 
$\rho=\mathcal{Z}^{-1}e^{-\sum_j\left(\frac{\epsilon_j}{k_BT}\hat{n}_j\right)}$, where the partition function is $\mathcal{Z}=Tr\left\{e^{-\sum_j\left(\frac{\epsilon_j}{k_BT}\hat{n}_j\right)}\right\}$. Hence, the probability of finding one photon at mode $i$ and its twin at mode $j$, namely $|c_{ij}|^2=Tr\left\{\rho \ket{1_i,1_j}\bra{1_i,1_j}\right\}$, is described by the Boltzmann distribution
\begin{align}\label{eq:CsTwoPh}
|c_{ij}|^2 = \frac{e^{\pm\epsilon_{ij}/T}}{\sum_{ij}e^{\pm\epsilon_{ij/T}}},
\end{align}
where \( T \) is the temperature, and we have set the Boltzmann constant to unity \( k_B = 1 \). Note, the plus sign $(+)$ in Eq.~\eqref{eq:CsTwoPh} accounts for the spectrum inversion of photonic waveguide systems, while $(-)$ corresponds to the 3D cavity.
In contrast to the single-photon scenario, the present results indicate that the thermalized two-photon states are described by the Boltzmann distribution. The predicted values for the temperature are \( T = 4.91 \) for the \( J_x \) lattice; \( T = -2.38 \) for the two-dimensional waveguide array; and \( T = 1.15 \) for the coupled cavity system. Notice that the distribution for the cavity accounts for the inverted eigenvalue spectrum. Furthermore, using $\rho$ one can show that $Tr\{\rho^2\} = \sum_mp_m^2\neq1$ which implies that the ensemble represents a fully mixed state. To confirm this, we evaluated the time-averaged elements of the corresponding density matrix and verified that the off-diagonal coherence terms decay to zero.\\
\indent
These findings are remarkable on several fronts. First, the thermal distributions describing two-photon states differ from those of the classical light field responsible for inducing nonlinear thermalization. Second, the optical temperature of the quantum state of light is not equal to that of the classical field, despite their direct interaction. In other words, thermal equilibrium is established at distinct effective optical temperatures for the classical and quantum subsystems. This discrepancy persists even when attempting to fit the quantum thermal state to the Rayleigh-Jeans distribution. 
According to the classical nonlinear thermalization theory \cite{Parto2019_Thermodynamic}, the value of the temperature $T$ 
depends on the internal energy $U$ and the total power $P$, which in turn depends on the mode amplitudes $c_i$ contained in the initial field (see definitions above). 
For classical light and single photons, the total number of accessible eigenmodes is $M$, as a result both obey the RJ distribution. On the other hand, for two-photon light there is a total of $M+\frac{M(M-1)}{2}$ accessible eigenmodes. This implies that the internal energy for two-photon light is given as $U=\sum_{m=1}^{M}\epsilon_{m,m}|c_{m,m}|^{2} +
\sum_{m=1}^{M}\sum_{n=m+1}^{M}\epsilon_{m,n}|c_{m,n}|^2$, where $\epsilon_{m,n}$ are the two-photon eigenvalues. That is, in the classical and quantum scenarios the constants defining temperature differ. Consequently, the optical temperature in the quantum case is different from the classical. 

\section{DISCUSSION}

In this work, we have demonstrated that the nonlinear thermalization of a classical field can drive quantum states of light toward thermal equilibrium. In the single-photon regime, the quantum state relaxes to a thermal distribution that closely matches that of the classical field. In contrast, for two-photon states, the resulting thermal distribution more closely resembles Boltzmann statistics, albeit with an effective optical temperature that differs from that of the classical field. While this intriguing discrepancy warrants further investigation, the physical mechanism underlying thermalization can be attributed to the chaotic dynamics of the classical field. Specifically, the results presented in the previous sections strongly suggest that nonlinear propagation of the classical light generates chaotic spatial fluctuations in the refractive index, which in turn induce thermalization of the quantum light.

These findings establish a direct connection between nonlinear optical systems and interacting many-body configurations, opening promising avenues for inducing a broader range of quantum effects within classical platforms. Importantly, a deeper understanding of nonlinearity-induced quantum state thermalization will enable the development of strategies to control this process, whether to accelerate, decelerate, or even suppress thermalization altogether. For example, studying the saturable Nonlinear Schrödinger Equation (NLSE), which admits analytical solutions, may provide valuable insights into engineering quantum light states that are robust against thermalization, akin to the concept of principal modes used to mitigate modal dispersion. 

\section{acknowledgments}
R.E. acknowledges support from the Army Research Office (W911NF23-1-0312) and AFOSR Multidisciplinary University Research Initiative Award on Programmable Systems with Non-Hermitian Quantum Dynamics (Grant No.FA9550-21-1-0202). A P.-L. is supported by MURI grant from Air Force Research Office (programmable systems with non-Hermitian quantum dynamics: FA9550-21-1-0202).

\bibliography{references.bib}

@Article{Wu2019_Thermodynamic,
  author    = {Wu, Fan O and Hassan, Absar U and Christodoulides, Demetrios N},
  journal   = {Nature Photonics},
  title     = {Thermodynamic theory of highly multimoded nonlinear optical systems},
  year      = {2019},
  number    = {11},
  pages     = {776--782},
  volume    = {13},
  publisher = {Nature Publishing Group},
}

@Article{Aschieri2011_Condensation,
  author    = {Aschieri, P and Garnier, Josselin and Michel, Claire and Doya, V and Picozzi, Antonio},
  journal   = {Physical Review A},
  title     = {Condensation and thermalization of classsical optical waves in a waveguide},
  year      = {2011},
  number    = {3},
  pages     = {033838},
  volume    = {83},
  publisher = {APS},
}

@Article{Picozzi2012_Condensation,
  author    = {Picozzi, Antonio and Rica, Sergio},
  journal   = {Optics Communications},
  title     = {Condensation of classical optical waves beyond the cubic nonlinear Schrödinger equation},
  year      = {2012},
  number    = {24},
  pages     = {5440--5448},
  volume    = {285},
  publisher = {Elsevier},
}

@Article{Pourbeyram2022_Direct,
  author    = {Pourbeyram, Hamed and Sidorenko, Pavel and Wu, Fan O and Bender, Nicholas and Wright, Logan and Christodoulides, Demetrios N and Wise, Frank},
  journal   = {Nature Physics},
  title     = {Direct observations of thermalization to a Rayleigh--Jeans distribution in multimode optical fibres},
  year      = {2022},
  number    = {6},
  pages     = {685--690},
  volume    = {18},
  publisher = {Nature Publishing Group UK London},
}

@Article{Marques2023_Observation,
  author    = {Marques Muniz, AL and Wu, FO and Jung, PS and Khajavikhan, M and Christodoulides, DN and Peschel, U},
  journal   = {Science},
  title     = {Observation of photon-photon thermodynamic processes under negative optical temperature conditions},
  year      = {2023},
  number    = {6636},
  pages     = {1019--1023},
  volume    = {379},
  publisher = {American Association for the Advancement of Science},
}

@Article{Parto2019_Thermodynamic,
  author    = {Parto, Midya and Wu, Fan O and Jung, Pawel S and Makris, Konstantinos and Christodoulides, Demetrios N},
  journal   = {Optics Letters},
  title     = {Thermodynamic conditions governing the optical temperature and chemical potential in nonlinear highly multimoded photonic systems},
  year      = {2019},
  number    = {16},
  pages     = {3936--3939},
  volume    = {44},
  publisher = {Optical Society of America},
}

@Article{Zhong2023_Universality,
  author    = {Zhong, Qi and Wu, Fan O and Hassan, Absar U and El-Ganainy, Ramy and Christodoulides, Demetrios N},
  journal   = {Nature Communications},
  title     = {Universality of light thermalization in multimoded nonlinear optical systems},
  year      = {2023},
  number    = {1},
  pages     = {370},
  volume    = {14},
  publisher = {Nature Publishing Group UK London},
}

@article{Peyronel2012,
	author = {Peyronel, Thibault and Firstenberg, Ofer and Liang, Qi-Yu and Hofferberth, Sebastian and Gorshkov, Alexey V. and Pohl, Thomas and Lukin, Mikhail D. and Vuletic, Vladan},
	date = {2012/08/01},
	doi = {10.1038/nature11361},
	id = {Peyronel2012},
	isbn = {1476-4687},
	journal = {Nature},
	number = {7409},
	pages = {57--60},
	title = {Quantum nonlinear optics with single photons enabled by strongly interacting atoms},
	volume = {488},
	year = {2012},
}

@article{Chang-Lukin2014,
	author = {Chang, Darrick E. and Vuletic, Vladan and Lukin, Mikhail D.},
	date = {2014/09/01},
	doi = {10.1038/nphoton.2014.192},
	id = {Chang2014},
	isbn = {1749-4893},
	journal = {Nature Photonics},
	number = {9},
	pages = {685--694},
	title = {Quantum nonlinear optics ---photon by photon},
	volume = {8},
	year = {2014},
	}

@Article{Perez2013_Coherent,
  author    = {Perez-Leija, Armando and Keil, Robert and Kay, Alastair and Moya-Cessa, Hector and Nolte, Stefan and Kwek, Leong-Chuan and Rodriguez-Lara, Blas M and Szameit, Alexander and Christodoulides, Demetrios N},
  journal   = {Physical Review A-Atomic, Molecular, and Optical Physics},
  title     = {Coherent quantum transport in photonic lattices},
  year      = {2013},
  number    = {1},
  pages     = {012309},
  volume    = {87},
  publisher = {APS},
}

@Article{Perez2013_Perfect,
  author    = {Perez-Leija, Armando and Keil, Robert and Moya-Cessa, Hector and Szameit, Alexander and Christodoulides, Demetrios N},
  journal   = {Physical Review A-Atomic, Molecular, and Optical Physics},
  title     = {Perfect transfer of path-entangled photons in J x photonic lattices},
  year      = {2013},
  number    = {2},
  pages     = {022303},
  volume    = {87},
  publisher = {APS},
}

@Article{Tschernig2018Multiphoton,
  author    = {Tschernig, Konrad and Leon-Montiel, Roberto de J and Magana-Loaiza, Omar S and Szameit, Alexander and Busch, Kurt and Perez-Leija, Armando},
  journal   = {J. Opt. Soc. Am. B},
  title     = {Multiphoton discrete fractional Fourier dynamics in waveguide beam splitters},
  year      = {2018},
  number    = {8},
  pages     = {1985--1989},
  volume    = {35},
  publisher = {Optica Publishing Group},
}

@Book{Loudon2000_The,
  author    = {Loudon, Rodney},
  publisher = {Oxford University Press},
  title     = {The quantum theory of light},
  year      = {2000},
  address   = {Oxford},
  edition   = {1st},
}

@article{Eberly2007,
author = {J. H. Eberly  and Ting Yu },
title = {The End of an Entanglement},
journal = {Science},
volume = {316},
number = {5824},
pages = {555-557},
year = {2007},
doi = {10.1126/science.1142654},
}

@article{PolkovnikovRevModPhys,
  title = {Colloquium: Nonequilibrium dynamics of closed interacting quantum systems},
  author = {Polkovnikov, Anatoli and Sengupta, Krishnendu and Silva, Alessandro and Vengalattore, Mukund},
  journal = {Rev. Mod. Phys.},
  volume = {83},
  issue = {3},
  pages = {863--883},
  numpages = {0},
  year = {2011},
  month = {Aug},
  publisher = {American Physical Society},
  doi = {10.1103/RevModPhys.83.863},
}

@article{LindenPRE,
  title = {Quantum mechanical evolution towards thermal equilibrium},
  author = {Linden, Noah and Popescu, Sandu and Short, Anthony J. and Winter, Andreas},
  journal = {Phys. Rev. E},
  volume = {79},
  issue = {6},
  pages = {061103},
  numpages = {12},
  year = {2009},
  month = {Jun},
  publisher = {American Physical Society},
  doi = {10.1103/PhysRevE.79.061103},
}

@Book{StrogatzBook,
  author    = {H. Strogatz, Steven},
  publisher = {CRS Press},
  title     = {Nonlinear dynamics and chaos},
  year      = {2018},
  address   = {},
  edition   = {2nd},
}

@Book{ChaikinBook,
  author    = { Chaikin, P. M.; Lubensky, T. C.},
  publisher = { Cambridge: Cambridge University Press},
  title     = {Principles of condensed matter physics },
  year      = {2007},
  address   = {},
  edition   = {2nd},
}

@article{Rigol2008,
	author = {Rigol, Marcos and Dunjko, Vanja and Olshanii, Maxim},
	date = {2008/04/01},
	doi = {10.1038/nature06838},
	id = {Rigol2008},
	isbn = {1476-4687},
	journal = {Nature},
	number = {7189},
	pages = {854--858},
	title = {Thermalization and its mechanism for generic isolated quantum systems},
	volume = {452},
	year = {2008}
	}

@article{RigolPRL2012,
  title = {Alternatives to Eigenstate Thermalization},
  author = {Rigol, Marcos and Srednicki, Mark},
  journal = {Phys. Rev. Lett.},
  volume = {108},
  issue = {11},
  pages = {110601},
  numpages = {5},
  year = {2012},
  month = {Mar},
  publisher = {American Physical Society},
  doi = {10.1103/PhysRevLett.108.110601}
}

@article{GoldsteinPRL2015,
  title = {Thermal Equilibrium of a Macroscopic Quantum System in a Pure State},
  author = {Goldstein, Sheldon and Huse, David A. and Lebowitz, Joel L. and Tumulka, Roderich},
  journal = {Phys. Rev. Lett.},
  volume = {115},
  issue = {10},
  pages = {100402},
  numpages = {5},
  year = {2015},
  month = {Sep},
  publisher = {American Physical Society},
  doi = {10.1103/PhysRevLett.115.100402}
}

@article{LevyPRB2018,
  title = {Equilibrium temperatures of discrete nonlinear systems},
  author = {Levy, Uri and Silberberg, Yaron},
  journal = {Phys. Rev. B},
  volume = {98},
  issue = {6},
  pages = {060303},
  numpages = {5},
  year = {2018},
  month = {Aug},
  publisher = {American Physical Society},
  doi = {10.1103/PhysRevB.98.060303}
}

@article{LevyJoPB2018,
doi = {10.1088/1361-6455/aa9a97},
year = {2018},
month = {jan},
publisher = {IOP Publishing},
volume = {51},
number = {3},
pages = {035401},
author = {Levy, Uri and Yang, Ken and Matzliah, Noam and Silberberg, Yaron},
title = {Universal correlations after thermalization in periodic nonlinear systems},
journal = {Journal of Physics B: Atomic, Molecular and Optical Physics}
}

@article{SilberbergPRL2009,
  title = {Universal Correlations in a Nonlinear Periodic 1D System},
  author = {Silberberg, Yaron and Lahini, Yoav and Bromberg, Yaron and Small, Eran and Morandotti, Roberto},
  journal = {Phys. Rev. Lett.},
  volume = {102},
  issue = {23},
  pages = {233904},
  numpages = {4},
  year = {2009},
  month = {Jun},
  publisher = {American Physical Society},
  doi = {10.1103/PhysRevLett.102.233904}
}

@article{BrenesPRLett2020,
  title = {Eigenstate Thermalization in a Locally Perturbed Integrable System},
  author = {Brenes, Marlon and LeBlond, Tyler and Goold, John and Rigol, Marcos},
  journal = {Phys. Rev. Lett.},
  volume = {125},
  issue = {7},
  pages = {070605},
  numpages = {6},
  year = {2020},
  month = {Aug},
  publisher = {American Physical Society},
  doi = {10.1103/PhysRevLett.125.070605}
}

@article{Vidmar2016,
doi = {10.1088/1742-5468/2016/06/064007},
year = {2016},
month = {jun},
publisher = {IOP Publishing and SISSA},
volume = {2016},
number = {6},
pages = {064007},
author = {Vidmar, Lev and Rigol, Marcos},
title = {Generalized Gibbs ensemble in integrable lattice models},
journal = {Journal of Statistical Mechanics: Theory and Experiment}
}

\end{document}